\documentclass[conference,compsoc]{IEEEtran}

\ifCLASSOPTIONcompsoc
  \usepackage[nocompress]{cite}
\else
  \usepackage{cite}
\fi

\usepackage{graphicx}
\usepackage{xspace}
\usepackage{enumitem}
\usepackage{hyperref}
\usepackage{xcolor}       
\usepackage{pifont}       
\usepackage{tikz}         
\usepackage{soul}
\sethlcolor{yellow}
\usepackage{amsmath}    
\usepackage{amsfonts}
\usepackage{booktabs} 
\usepackage[capitalize,noabbrev]{cleveref}
\usepackage{amsthm}
\usepackage{multirow}
\usepackage{multicol}
\usepackage{booktabs}
\theoremstyle{definition} 
 
\usepackage{tabularx}
\usepackage{array}
\usepackage{colortbl}
\usepackage[normalem]{ulem}
\usepackage{xcolor}

\usepackage{xcolor}
\usepackage[normalem]{ulem}

\newif\ifrevised
\revisedfalse

\ifrevised
  \newcommand{\added}[1]{\textcolor{blue}{#1}}
  \newcommand{\addedd}[1]{\textcolor{green!60!black}{#1}}
  \newcommand{\deleted}[1]{\textcolor{red}{\sout{#1}}}
\else
  \newcommand{\added}[1]{#1}
  \newcommand{\addedd}[1]{#1}
  \newcommand{\deleted}[1]{}
\fi

\newcounter{note}[section]

\newcommand{\bheading}[1]{\vspace{2pt}\noindent\textbf{#1}}

\newcounter{insights}

\setlist[itemize]{leftmargin=*}
\setlist[enumerate]{leftmargin=*}

\newcommand{\secref}[1]{\mbox{Section~\ref{#1}}\xspace}

\newcommand{\figref}[1]{\mbox{Figure~\ref{#1}}\xspace}
\newcommand{\tabref}[1]{\mbox{Table~\ref{#1}}\xspace}

\newcounter{packednmbr}

\newenvironment{packeditemize}{
\begin{list}{$\bullet$}{
\setlength{\labelwidth}{0pt}
\setlength{\itemsep}{0pt}
\setlength{\leftmargin}{\labelwidth}
\addtolength{\leftmargin}{\labelsep}
\setlength{\parindent}{0pt}
\setlength{\listparindent}{\parindent}
\setlength{\parsep}{0pt}
\setlength{\topsep}{0pt}}}{\end{list}}

\begin{document}

\pagenumbering{gobble}

\title{Hollow-LLM Attack: Computationally Trivial Weights in Zero-Knowledge Verification of LLM Inference}
\author{
    \IEEEauthorblockN{Chen Gong}
    \IEEEauthorblockA{
        University of Southern California\\
        cgong459@usc.edu
    }
    \and
    \IEEEauthorblockN{Beijie Liu}
    \IEEEauthorblockA{
        University of Southern California\\
        beijieli@usc.edu
    }
    \and
    \IEEEauthorblockN{Mengyuan Li}
    \IEEEauthorblockA{
        University of Southern California\\
        mli49061@usc.edu
    }
}
\maketitle
\begingroup
\renewcommand{\thefootnote}{}
\footnotetext{\copyright~2026 IEEE. Personal use of this material is permitted. Permission from IEEE must be obtained for all other uses, in any current or future media, including reprinting/republishing this material for advertising or promotional purposes, creating new collective works, for resale or redistribution to servers or lists, or reuse of any copyrighted component of this work in other works.}
\endgroup
\thispagestyle{plain}
\pagestyle{plain}

\IEEEpeerreviewmaketitle

\begin{abstract}
As large language models (LLMs) grow in scale and are predominantly served from remote platforms, verifying faithful inference execution becomes critical (i.e., ensuring that a provider actually executes the advertised model and computational workload rather than a tampered or downsized variant). Zero-knowledge (ZK) LLM inference offers an appealing approach. It promises public verifiability and delivers per-instance guarantees of equational correctness by proving that an output is consistent with executing a public architecture under committed, private weights. Though, we show that it does not bind the effort expended to produce the output.

In this paper, we formalize this overlooked effort gap and introduce the Hollow-LLM Attack, in which a dishonest provider retains the declared architecture and parameter count but embeds ghost weights whose algebraic structure collapses effective computation. These witnesses satisfy the verification circuit and yield valid proofs, even though the dishonest model owner, who serves as the prover, performs computation commensurate with a much smaller model than the declared public architecture. This creates a profitable equilibrium in which providers deliver provably correct outputs at small-model cost while overclaiming model size. Accordingly, we characterize concrete families of ghost weights that compose with standard transformer blocks and show that such hollow deployments substantially reduce serving cost with zero quality loss under the same verification circuit. These findings underscore that proof of correct inference is not proof of large-model execution and necessitate additional protections to bind correctness to verifiable computational work.
\end{abstract}
\section{Introduction}
\label{sec:introduction}

Large Language Models (LLMs) have become increasingly powerful and pervasive, underpinning a large market of AI services. With the growth in both size and scale, LLMs usually require substantial GPU resources and are thus often deployed on remote or cloud  platforms. This deployment model raises critical concerns about the \emph{faithful execution} of inference: \textit{how can users trust that an LLM service is faithfully running the claimed model?} Such concerns include verifying that the deployed model has not been tampered with (e.g., no hidden backdoors)~\cite{ge2025backdoors, li2024backdoorllm, li2021hidden}, that the outputs are truly generated by the advertised model~\cite{sun2025coin, chen2025zktorch, sun:2024:zkllm, qu:2025:zkgpt, pasquini2025llmmap, wang2022ezdps, chen2024zkml}, and that the model’s size or parameters have not been surreptitiously replaced with a smaller or different model~\cite{sun2024svip, cai2025you, zhu2025auditing, yuan2025trustverify, gao2024model, alhazbi2025llms}.

To address this trust issue, the community has explored various privacy-preserving verification techniques. Approaches such as confidential computing (trusted execution environments)~\cite{tramer2018slalom, hunt2018chiron, nvidia:2023:h100}, homomorphic encryption~\cite{gilad2016cryptonets, lee2022privacy, lee2023precise, jin2023fedml, effendi2024privacy}, and zero-knowledge proofs (ZKPs)~\cite{wang2022ezdps, chen2024zkml, sun:2024:zkllm, qu:2025:zkgpt, chen2025zktorch}, have been proposed to enforce that an LLM inference was performed faithfully by the intended model without revealing proprietary model weights. Among these, zero-knowledge machine learning (zkML) stands out as a promising direction for publicly verifiable inference.
A typical zkML workflow for LLM inference proceeds as follows: (i) the provider makes public the model architecture and a \emph{commitment} to the \addedd{proprietary private} model weights (e.g., via a cryptographic hash); (ii) the user sends a prompt to the provider; (iii) the prover (i.e., the provider) generates a succinct cryptographic proof asserting that the produced output results from executing the declared model on the given prompt; and (iv) the verifier (i.e., the user) checks this proof. \deleted{Crucially, the proof hides the actual weight values while guaranteeing that they \emph{exist} and produce the result.} \added{Crucially, the proof hides the actual weight values while guaranteeing that the output is consistent with committed private weights under the declared architecture. This commitment-only design is especially appealing for proprietary ML services because it offers public verifiability 
without revealing weights or relying on a trusted third party.} 

Recent systems demonstrate the feasibility of end-to-end ZK LLM inference even for multi-billion-parameter models, yielding compact proofs in practical time. For example, zkLLM~\cite{sun:2024:zkllm} generates a proof for a 13B-parameter transformer in under 15 minutes with proof size around 200~KB, preserving model-weight privacy. Similarly, zkGPT~\cite{qu:2025:zkgpt} proves GPT-2 inference in about 25 seconds by introducing optimized constraints for transformer layers. These advances show that ZKP can ensure users of correct inference by large language models, instilling confidence that the model output was indeed correct inference results from the declared model.

In this paper, we reveal a previously overlooked limitation of the above ZK LLM inference verification procedure. Specifically, a zero-knowledge proof of inference certifies membership in a nondeterministic polynomial-time (NP) relation: it proves that there exist private weights such that the output is the result of executing the public model architecture on the input. However, the proof does not attest to the algorithmic path taken to obtain that result or how much computation was necessary to get the result. 
In other words, it verifies the correctness of the equations that define the model size and architecture, but not the effort expended to calculate the result. This creates an \textbf{effort gap} compared to the declared public model size: \textit{a dishonest provider can furnish a trivially valid witness that yields a verifiable output while performing far less work than the declared public architecture suggests.} 
\added{We emphasize that this issue is not a flaw in ZKP soundness. Rather, it is a deployment-level assurance gap: the proof certifies consistency with a committed witness under the declared computation, but that guarantee does not inherently bind the amount of computation actually expended.}

We concretize this gap through the \emph{Hollow-LLM attack}. Our setting follows existing zero-knowledge LLM-inference deployments, such as zkGPT~\cite{qu:2025:zkgpt} and zkLLM~\cite{sun:2024:zkllm}, where the architecture is declared public, the weights are private but cryptographically committed, and the proofs accompany inference or periodic audits. Within this setting, a dishonest provider can deploy a \emph{hollow LLM}: the provider publicly declares a larger model (the \textbf{\textit{Outer Model}}), then carefully fills the larger architecture with a smaller model augmented by \emph{ghost weights}, and finally commits the entire hollow model. These ghost weights are chosen so that the declared layers either perform negligible computation or reuse the computation of the much smaller \textbf{\textit{Inner Model}}.
During service, the provider computes outputs using only the smaller inner model. However, because the ghost weights are crafted appropriately, the provider can still supply a proof asserting that the output was computed by the declared larger outer model, under the committed weights and on the same input prompt.
In this hollow LLM deployment, all public interfaces remain intact. The layer layout, tensor shapes, and stated parameter count all match the declared model, so both the proof and associated metadata appear legitimate. The deviation lies entirely in the private algebraic structure of the ghost weights: some layers effectively pass their inputs through unchanged, while wide layers are arranged so that only a small subspace carries the actual signal and the remainder remains inactive. Consequently, the computation follows that of the smaller inner model while the verifier continues to endorse the claimed outer model. Formal definitions and concrete constructions appear in \secref{sec:framework} and \secref{sec:attack}.

\begin{figure}[t]
\centering
\includegraphics[width=\columnwidth]{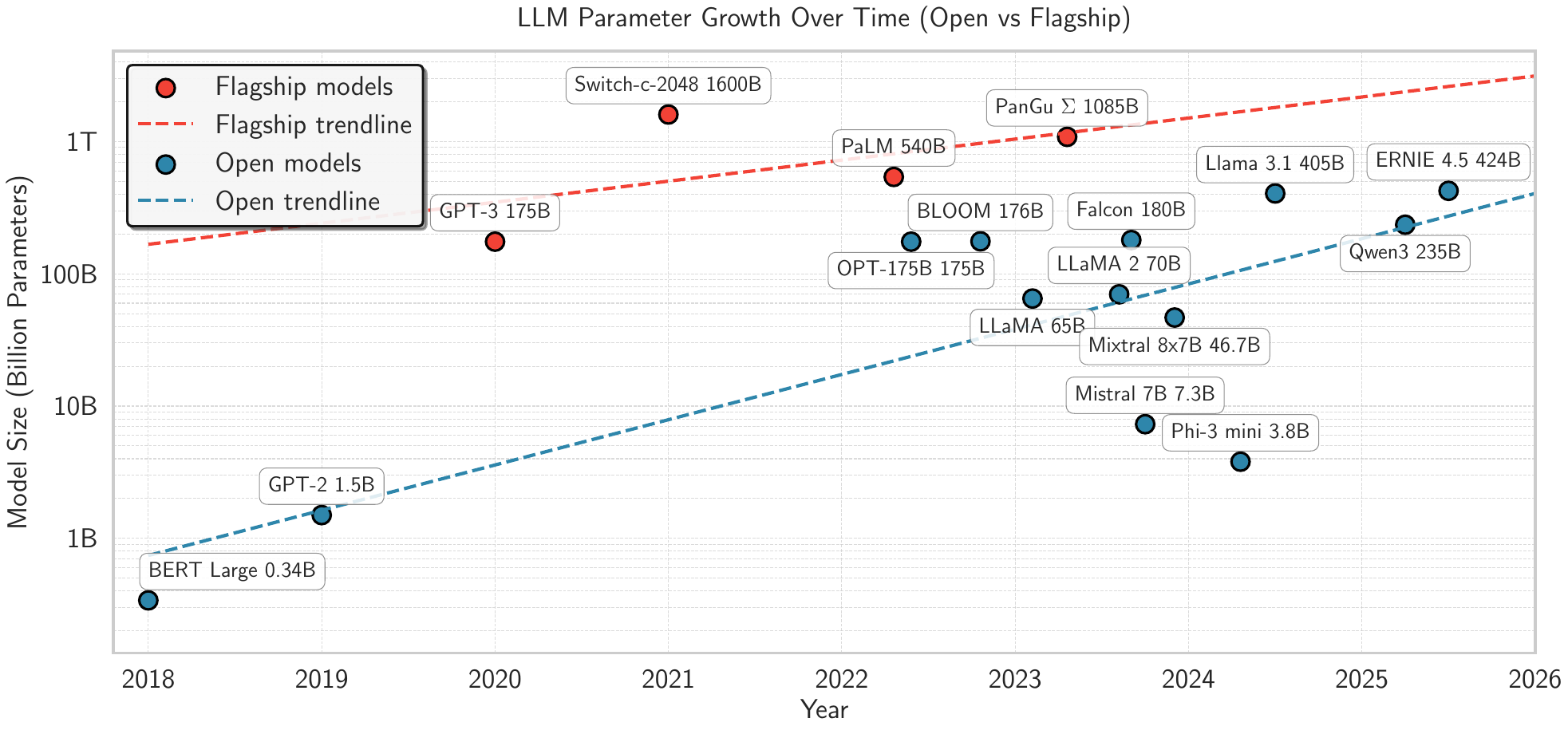}
\vspace{-15pt}
\caption{Growth trend of LLM parameters over time, distinguishing open-source and flagship models.}
\label{fig:size_grow}
\end{figure}

We consider the Hollow-LLM attack particularly concerning for several reasons: 
\textbf{(1) Incentive to exaggerate the scale.} In the release and service of the LLM model, the number of parameters remains the most visible indicator of the capability and is continuously increasing; and companies prominently advertise the size of the model during public releases (the trend is illustrated in \figref{fig:size_grow}). Previous studies also show that users tend to associate larger models and longer results with greater precision and reliability~\cite{steyvers:2025:large, lehmann:2022:risk,mcconnon:2024:bigger}, further motivating providers to overstate the scale of their models. 
\textbf{(2) Resource-linked incentives.} Many pricing and governance schemes tie resource quotas and costs directly to the declared model size. Inflating the nominal parameter count can, therefore, yield both reputational and financial advantages. 
\deleted{\textbf{(3) Stealthiness.} }
\added{\textbf{(3) Proof-level indistinguishability}}
\deleted{The Hollow-LLM attack is entirely covert.} 
    \added{At the proof level, Hollow-LLM is covert: accepted transcripts remain consistent with the declared architecture and committed weights. Neither party can distinguish a hollowed model from a genuine one by comparing zero-knowledge proofs alone, since both produce proofs consistent with the public architecture.}
Existing clients and verifiers cannot \textit{reliably} and \textit{directly} infer the true model size because the model performance does not scale linearly with the parameter count~\cite{gao2024model,cai2025you,hinton2015distilling, sanh2019distilbert}. 
\added{Deployment-level audits such as behavioral checks, reference-based comparisons, or telemetry may still detect aggressive over-claiming, but evaluating such detection signals is beyond the scope of this paper. Our focus here is the proof-level gap.}
\textbf{(4) Cost reduction.} A dishonest provider can generate outputs that satisfy the ZK verification circuit and produce valid proofs while performing only a fraction of the computation implied by the declared model.

Notably, we observe that this kind of misdirection appears to be underappreciated in the emerging zkML community. several major blockchain and crypto infrastructure sources present ZK LLM inference in ways that implicitly equate proof generation with computational effort. For example, the Ledger Academy glossary, maintained by Ledger, one of the largest global crypto hardware wallet manufacturers, states that a ZK LLM inference “certificate contains details such as model size and parameters, confirming that a certain computation has been done”~\cite{ledger:ZkML:2025}. Likewise, CoinGecko, a leading cryptocurrency market data provider, defines ZK LLM as a system that verifies inference “including details like model size and parameters”~\cite{coingecko:ZkML:2024}. 
These descriptions reveal a widespread misconception that ZK proofs of LLM inferences inherently attest to computational scale, highlighting why the Hollow-LLM attack and the resulting effort gap constitute a pressing and emerging concern.


The contributions of this paper are summarized as:

\begin{packeditemize}

\item We identify and formalize an \emph{effort gap} in current ZK-verified LLM inference: existing schemes certify membership in an NP relation but provide no guarantee that the prover expends computation commensurate with the advertised model size, exposing a mismatch between what ZK proofs attest to and the assurances assumed in LLM service markets.

\item We introduce the Hollow-LLM attack model, in which a dishonest provider publicly declares a large outer model while privately deploying a smaller inner model. Using \emph{ghost weights}, the provider preserves the public architecture, parameter count, and proof validity while routing almost all computation through the inner model, leaving verifiers unable to distinguish a hollowed deployment from a genuine one.

\item We develop two training-free algebraic constructions of hollow LLMs that are compatible with state-of-the-art zkLLM pipelines. These constructions exploit transformer invariances to insert effectively identity layers and to inflate embedding dimensions while confining activations to a low-dimensional subspace, thereby decoupling publicly claimed depth and width from the true per-token computational cost.

\item We analyze the security and economic implications of the effort gap in realistic deployment scenarios, showing that a provider can pass ZK verification while using only a fraction of the computation implied by the declared model. This creates strong incentives to exaggerate model scale and challenges emerging zkML narratives that equate proof validity with faithful execution effort.

\end{packeditemize}





\section{Background}

\subsection{Transformer Architecture}
\label{sec:transformer-arch}

\begin{figure}[t]
  \centering
  \includegraphics[width=\linewidth]{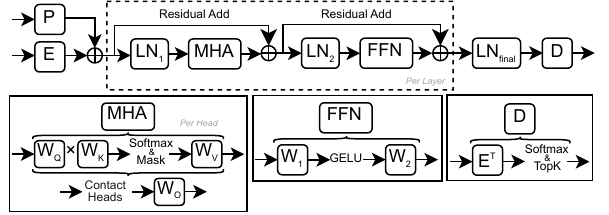}
  \caption{Orientation diagram for the pre-LN transformer: embedding $E$, positional $P$, attention $(W_Q,W_K,W_V,W_O)$, feed-forward $(W_1,W_2)$, and decoder $D$. Components outside the proof boundary are system dependent and stated when relevant.}
  \label{fig:skeleton}
\end{figure}

In this paper, we focus on the standard decoder-only Transformer architecture used in contemporary LLMs and inference serving. 
\figref{fig:skeleton} illustrates a single pre-LN Transformer block, and \tabref{tab:symbols} summarizes the notation for both the base architecture and the additional symbols introduced later for Hollow-LLM attacks. 
Here we briefly fix conventions for a single layer.

Let $x_{1:T}$ denote an input token sequence of length $T$, and let 
$X \in \mathbb{R}^{T \times d_{\text{model}}}$ be the corresponding stack of hidden states after token embedding and positional encoding.
We adopt the usual factorization $d_{\text{model}} = h\,d_k = h\,d_v$ across $h$ attention heads (see \tabref{tab:symbols}), and write the layer computation in pre-LN form as
\[
\begin{aligned}
&\tilde{X} = \mathrm{LN}(X), \quad
Q = \tilde{X} W_Q,\;
K = \tilde{X} W_K,\;
V_{\text{att}} = \tilde{X} W_V,\\
&H_{\text{att}} = \mathrm{softmax}\!\Bigl(\tfrac{1}{\sqrt{d_k}} Q K^\top\Bigr) V_{\text{att}}, \quad
X^{\mathrm{att}} = X + H_{\text{att}} W_O,\\
&\tilde{X}' = \mathrm{LN}(X^{\mathrm{att}}), \quad
X' = X^{\mathrm{att}} + \varphi(\tilde{X}' W_1)\,W_2,
\end{aligned}
\]
followed by a final layer normalization $\mathrm{LN}_{\mathrm{final}}$.
Here $\mathrm{LN}$ denotes a learned LayerNorm applied row-wise, $W_Q, W_K, W_V, W_O$ are the attention projection matrices, $W_1, W_2$ are the feedforward weights, and $\varphi$ is the pointwise nonlinearity (e.g., GELU).

The input pipeline maps discrete tokens in the vocabulary to rows of $X$ via an embedding matrix $E$ together with a positional mechanism $P$.
The output pipeline maps the final hidden state to logits via a decoder matrix $D$, which may be tied or untied with $E$ depending on the deployment.


\subsection{Zero-knowledge LLM inference}
\label{sec:background-zkllm}

\begin{figure}[t]
  \centering
  \includegraphics[width=\linewidth,page=1]{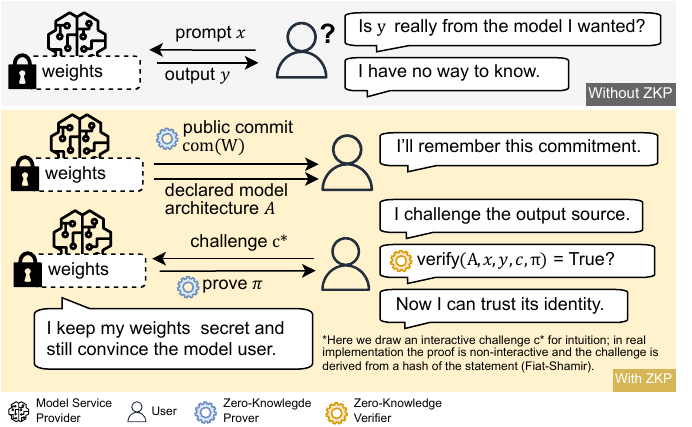}
  \caption{Ordinary serving versus zero-knowledge serving.
The provider commits to private weights $W$ and returns $(y,\pi)$;
the verifier checks that the declared architecture $A$, with some opening of $\mathsf{Com}(W)$, produces $y$ on input $x$ without revealing $W$.}
  \label{fig:zkllm-before-after}
\end{figure}

Figure~\ref{fig:zkllm-before-after} contrasts ordinary remote LLM serving
with a zero-knowledge deployment~\cite{qu:2025:zkgpt,sun:2024:zkllm}.
A user sends a prompt $x$ to a service provider and receives an output
$y$.
The model architecture $A$ (layer layout, widths, heads, positional
encoding, normalization scheme, and tying choices) is public, while the
weights $W$ remain private.
The provider publishes a binding commitment $\mathsf{Com}(W)$ to these
weights and, for each query, returns both the output $y$ and a proof~$\pi$.

\added{In this setting, the provider acts as the \emph{prover}, while the user
or auditor acts as the \emph{verifier}.
We use \emph{inference execution} to mean the actual forward-pass
computation performed by the provider on input $x$.
We say the service performs \emph{faithful inference} when the returned
output is exactly the result of running the declared architecture $A$
with the committed private weights $W$ on the same input.
The public \emph{statement} consists of $(A, x, y, \mathsf{Com}(W))$,
while the private \emph{witness} consists of the opening of the
commitment and, depending on the proving system, any auxiliary values
needed to satisfy the circuit.
}

The verifier runs
\[
\mathsf{Verify}\bigl(A, x, y, \mathsf{Com}(W), \pi\bigr)
  \in \{\mathsf{accept},\mathsf{reject}\}
\]
and accepts exactly when there exist weights $W$ that open the commitment
and make the declared architecture $A$ produce $y$ on input $x$.
In practice, an underlying interactive protocol is made non-interactive
via the standard Fiat--Shamir transform, so the verifier simply checks a
single proof $\pi$ for the public statement $(A,x,y,\mathsf{Com}(W))$.

Overall, the proof certifies that the served output is \emph{consistent}
with some fixed private weights under the public architecture, without
revealing those weights. \added{This guarantee is commitment-relative: it ties the output to the
committed private weights under the public architecture, rather than to
any external reference model or public weight hash.
Equivalently, acceptance establishes equation-level correctness of the
claimed inference relative to the declared statement, but does not by
itself certify how much computation or execution effort was expended to
produce that output.}

\newcolumntype{Y}{>{\raggedright\arraybackslash}X}
\newcolumntype{R}[1]{>{\raggedleft\arraybackslash$}m{#1}<{$}} 

\arrayrulecolor{black}
\setlength{\heavyrulewidth}{1pt}
\setlength{\lightrulewidth}{0.6pt}
\setlength{\cmidrulewidth}{\lightrulewidth}
\setlength{\aboverulesep}{0.3ex}
\setlength{\belowrulesep}{0.3ex}

\renewcommand{\arraystretch}{1.12}
\setlength{\tabcolsep}{4.5pt}

\begin{table}[t]
\centering
\small
\caption{\textbf{Symbols \& notation.} Core transformer symbols and attack-specific symbols.}
\label{tab:symbols}
\vspace{-5pt}
\begin{minipage}{0.48\textwidth}
\centering
\textsc{Group I: Symbols used in Transformer}\par\vspace{2pt}  
\begin{tabularx}{\linewidth}{@{}R{0.39\linewidth} Y@{}}
\toprule
\textbf{Symbol} & \textbf{Meaning} \\
\midrule
x_{1:T},\,T & input token sequence and its length \\
y\ \text{(or }y_{1:K}\text{)} & next-token output (or $K$-step decode) \\
V & vocabulary size \\
d_{\mathrm{model}},\,h,\,d_k,\,d_v,\,d_{\mathrm{ff}},\,L & model width, heads, per-head dims, FFN width, depth \\
E,\,D & token embedding and decoder (tied unless noted) \\
P & positional encoding (absolute or rotary) \\
W_Q,\,W_K,\,W_V,\,W_O & attention projection matrices \\
W_1,\,W_2;\ \varphi & FFN matrices; $\varphi$ nonlinearity \\
\mathrm{LN}_{\gamma,\beta} & LayerNorm with $\gamma,\beta\in\mathbb{R}^{d_{\mathrm{model}}}$ \\
\bottomrule
\end{tabularx}
\end{minipage}\hfill
\begin{minipage}{0.48\textwidth}
\centering\par\vspace{4pt}
\textsc{Group II: Symbols used in the  Attack}\par\vspace{2pt}
\begin{tabularx}{\linewidth}{@{}R{0.34\linewidth} Y@{}}
\toprule
\textbf{Symbol} & \textbf{Meaning} \\
\midrule
A_{\mathrm{out}} & declared public architecture checked by the verifier \\
A_{\mathrm{in}} & served inner architecture executed by the prover \\
W^{\mathrm{ghost}}_{\mathrm{out}} & ghost witness weights under $A_{\mathrm{out}}$ \\
W_{\mathrm{in}} & private weights for $A_{\mathrm{in}}$ \\
m & replication factor (Attack~B) \\
d'_{\mathrm{model}} & expanded width $m\,d_{\mathrm{model}}$ \\
R=\mathbf{1}_m\otimes I_{d_{\mathrm{model}}} & replication from $d_{\mathrm{model}}$ to $m d_{\mathrm{model}}$ \\
S_s & selector for block $s\in[m]$ \\
\mathrm{blkdiag}(M,\ldots,M) & block-diagonal with $m$ copies of $M$ \\
\bottomrule
\end{tabularx}
\end{minipage}

\end{table}


\subsection{Easy Witnesses in LLM Serving}

Work on verifiable computation has long noted the possibility of
``easy witnesses'': witnesses that satisfy all circuit constraints while
corresponding to much less computation than a naive effort model would
suggest~\cite{murray:2018:circuit,impagliazzo:2002:search,kattis:2023:proof}.
\added{In zk-LLM serving, this appears as a deployment-level effort-binding
gap: the proof certifies consistency with the declared circuit and committed
weights, but does not inherently bind the prover to the amount of
computation users may associate with the declared model size.}

\deleted{To our knowledge, no prior work instantiates easy witnesses as a concrete \emph{effort-binding} vulnerability for ZK-based LLM serving, and no existing ZK-based LLM system attempts to prevent or detect low-effort witnesses for a large declared model.}
\added{To our knowledge, prior work has not concretely instantiated this gap
for transformer-based zk-LLM serving. We show that transformer
architectures admit structured easy witnesses, realized as \emph{ghost
weights}, that preserve the public architecture and proof validity while
collapsing serve-time computation to that of a much smaller inner model.}




\section{Hollow-LLM Overview}
\label{sec:framework}

A Hollow-LLM attack is a non-faithful LLM inference serving deployment in which a provider serves queries using a smaller inner model $A_{\mathrm{in}}$, while supplying carefully chosen ghost weights so that the verifier accepts the transcript as if it were produced by the declared outer architecture $A_{\mathrm{out}}$ under the zk-LLM proof contract. In this section, we formalize the threat model, the attack objectives, and the target LLM architecture used in the hollow LLM deployment.

\subsection{Threat Model}
\label{sec:atk:threat}


\deleted{We consider a publicly verifiable zkML deployment for LLM inference (e.g., zkLLM~\cite{sun:2024:zkllm}, zkGPT~\cite{qu:2025:zkgpt}). }
\added{We consider proprietary-weight zkLLM inference deployments, consistent with zkLLM~\cite{sun:2024:zkllm} and zkGPT~\cite{qu:2025:zkgpt}.}
The declared architecture \(A_{\text{out}}\) (layer topology, widths, heads, positional encoding, normalization layout, and any parameter tying) is public and compiled into the verifier’s circuit. 
The model owner keeps weights \((W_{\text{out}})\) private and either commits to them in advance or supplies openings consistent with a prior commitment. 
We refer to the declared pair \((A_{\text{out}}, W_{\text{out}})\) as the \textit{\textbf{Outer model}}. 
Given an input prompt \(x\), the provider returns an output \(y\) and a zero-knowledge proof \(\pi\) that the circuit’s forward pass on \((A_{\text{out}}, W_{\text{out}}, x)\) yields \(y\), revealing nothing about the weights.
\added{Public-weight or reference-bound deployments are therefore out of scope.}

\bheading{Arithmetic and Determinism.}
The circuit implements a deterministic transformer forward pass using fixed-point/integer arithmetic with public rounding rules; nonlinearities (e.g., softmax, GELU) are realized via public lookup tables. 
Stochastic decoding, if used, is made reproducible by a public, committed seed that fixes the sampling trajectory. 
This ensures that the same \((x, W)\) and seed deterministically yield the same \(y\).
Quantization is a common implementation choice in most verifiable ML to match verifier arithmetic and avoid floating-point nondeterminism; the hollowing constructions themselves are algebraic and apply equally to floating-point models so long as prover and circuit share the same numeric semantics.

\bheading{Inference Serving vs.\ Proof Regime.}
Inference is served token-by-token, whereas proofs are generated per query transcript. In practice, inference results may be returned without real-time proof for non-interactive ZK proofs (~\cite{qu:2025:zkgpt}): when a specific call is selected for audit, the provider can reconstruct a witness for the already-served transcript (by re-running the deterministic computation or using cached intermediates) and produces a proof~$\pi$ off the critical path. This regime avoids the overhead of attaching a ZK proof to every response, while still enabling post-hoc audits that deter blatant cheating.

\bheading{Adversary Capabilities.}
The adversary is a dishonest service provider who seeks to reduce computation while appearing to run the larger declared model. 
Concretely, the provider may generate \(y\) using a smaller \emph{\textbf{Inner Model}} \((A_{\text{in}}, W_{\text{in}})\) at serve time, while convincing the verifier that \(y\) came from \(A_{\text{out}}\). 
The attacker selects \emph{ghost weights} \((W_{\text{out}}^{\text{ghost}})\) so that the circuit accepts \((x, y)\) as consistent with \(A_{\text{out}}\), while \(y\) can in fact be produced at the cost of executing only the smaller model \(A_{\text{in}}\).
Typically \(A_{\text{out}}\) strictly expands \(A_{\text{in}}\) (e.g., deeper or wider), enabling algebraic alignment between the outer equations and a cheaper computation path. 
The provider controls the online service and also acts as the prover, including batching, caching, scheduling, and persistence of intermediate tensors across serve, generating proof, as well as participating during the audit phases. 
\deleted{Depending on the commitment mode, weights may be fixed once (with subsequent openings) or chosen per call subject to binding commitments. }
\added{The commitment is binding across serving and audit; the attacker’s freedom lies in choosing the committed weights within this private-model regime, not in changing them after the fact.}
The adversary cannot modify the verifier’s circuit, trapdoors, or keys, and learns nothing beyond what the zero-knowledge procedure reveals.

\bheading{Verifier View.}
For each verification, the verifier (or the service client) observes only \((x, y, \pi)\) and an accept/reject bit. 
There is no trusted hardware attestation bound to runtime behavior, no external metering of FLOPs/energy, no side-channel telemetry, and no watermarking or behavioral fingerprinting assumed. 
Audits may be sparse and asynchronous relative to serving, reflecting practical deployments where proving is performed on demand or periodically rather than on every token. 
Verification runs the fixed circuit for \(A_{\text{out}}\) and is lightweight relative to proving (\secref{sec:experiments}).

\bheading{Out of scope.}
We exclude training provenance and dataset integrity, key management, network-layer timing attacks, and defenses that alter the circuit to measure work or introduce orthogonal trust anchors (e.g., TEEs with attestation). 
Our focus is the gap between \emph{equation-level correctness} and \emph{expended effort} in conventional ZK LLM workflows. 
The Hollow-LLM deployments instantiated later exploit these degrees of freedom and the audit-based operation to shift cost off the critical path, aiming at the goals in \secref{sec:atk:objective}.

\subsection{Attack Objectives}
\label{sec:atk:objective}


The main objective of the Hollow-LLM attack is to make a dishonest provider appear fully compliant with the verifier, while the provider can pay only the computational cost of a smaller model during inferencing. 
To match practical hollow LLM deployments, we target five concrete attack objectives:
\begin{itemize}\itemsep2pt
  \item \emph{\textbf{G1.} Contract preservation.} We do not modify the verifier, its circuit, the statement format, or public parameters (CRS, keys, API). Each proof still asserts that $y$ equals the forward pass of the declared architecture on $x$.
  
  \item \emph{\textbf{G2.} Asymmetric serving and proving.} Online inference generation uses a smaller inner model $(A_{\mathrm{in}},W_{\mathrm{in}})$, while the audit-time proof is issued against $A_{\mathrm{out}}$ using private \emph{ghost weights} $W_{\mathrm{out}}^{\mathrm{ghost}}$.
  
  \item \emph{\textbf{G3.} Serve-time efficiency.} Per-token latency and FLOPs at serve time should match $A_{\mathrm{in}}$, not $A_{\mathrm{out}}$; in practice, \deleted{the provider can always add artificial delay so that the verifier cannot distinguish hollowed runs from honest ones.} \added{the provider can add artificial delay, so raw response time alone is not a reliable distinguisher.}
  
  \item \emph{\textbf{G4.} Proof indistinguishability.}   For fixed $(A_{\mathrm{out}}, x, y)$ and a fixed circuit, accepted proofs from hollowed runs are distributed the same way as those from an honest run of $A_{\mathrm{out}}$.
  
  \item \emph{\textbf{G5.} Cost asymmetry.} Any extra effort is shifted off the latency path to the audit/proving phase (one proof per transcript), while the verification cost remains small as in standard zkML.
\end{itemize}

\subsection{General Attack Steps}
We outline the general Hollow-LLM attack steps below:

\begin{enumerate}
\item Declared Public Architecture and Committed Private Weights. 
The model provider originally has an inner model ($A_{\mathrm{in}}$, $W_{\mathrm{in}}$). The provider then privately chooses an appropriate size $A_{\mathrm{out}}$ and attack-specific \emph{ghost weights} $W_{\mathrm{out}}^{\mathrm{ghost}}$.The provider then declares the model architecture and commits $W_{\mathrm{out}}$.
With prompt input $(x)$, a set of committed weights, and a public fixed or derived decoding seed $s$, the later LLM inference serving service can be verified by ZKP.

\item  LLM Serving (per token, fast). The provider runs the inner model and returns tokens to the user, yielding
\[
  y_{\mathrm{in}} \;=\; f_{A_{\mathrm{in}}}(W_{\mathrm{in}}, x; s).
\]
\deleted{This phase intentionally hides user-visible latency, which therefore follows expected latency from $A_{\mathrm{out}}$.}
\added{This phase executes only $A_{\mathrm{in}}$ at serve time. Wall-clock latency may be padded, so raw response time alone is not a definitive distinguisher.}

\item ZKP Audit (per transcript, off the critical path). The user (i.e., verifier) can periodically choose a pair $(x, y_{\mathrm{in}})$ and ask the model provider for a ZK proof that the previous response was generated by $(A_{\mathrm{out}}, W_{\mathrm{out}})$. 
The attack-specific ghost weights $W_{\mathrm{out}}^{\mathrm{ghost}}$ must be carefully crafted in STEP 1 for the declared architecture $A_{\mathrm{out}}$ so that evaluating the verifier’s fixed circuit on the same $(x, s)$ reproduces the already served transcript. This should not be a one-shot coincidence, but should hold for any $(x, s)$ pair, since any change in $W_{\mathrm{out}}$ would be reflected in the commitment and detected by the user
:
\[
  f_{A_{\mathrm{out}}}(W_{\mathrm{out}}^{\mathrm{ghost}}, x; s) \;=\; y_{\mathrm{in}}.
\]

\item Between Serving and Audit (optional caching). The provider may optionally cache minimal state (e.g., the public seed, prompts, or selected intermediates) to speed up the auditing procedure. This is not required and is not visible to the verifier.

\item  Produce the Proof. Per audit request, the provider generates a proof $\pi$ using standard zkML machinery for the \emph{outer} circuit on $(x,y_{\mathrm{in}})$, and the verifier checks it exactly as in an honest run.
\end{enumerate}


\bheading{Success Criteria.}
We regard the attack as successful only if it simultaneously achieves correctness and the desired cost asymmetry. We focus on per-serving cost, since serving happens token by token and determines the model provider's cost of the service, whereas proving happens only on demand: the cost of generating a ZKP for LLM inference is order-of-magnitude times that of inference serving and can be scheduled asynchronously (21.8s for GPT-2 using 32 threads~\cite{qu:2025:zkgpt}). By construction,
\[
  T^{\mathrm{serve}}_{\mathrm{hollow}} \;=\; T^{\mathrm{serve}}(A_{\mathrm{in}}) \quad\ll\quad T^{\mathrm{serve}}(A_{\mathrm{out}}),
\]
so the provider saves serve-time cost. On an audited call, we regard the attack as successful when all criteria during verification hold:

\begin{itemize}\itemsep2pt
  \item \emph{\textbf{S1.} Accepted Proof.} The verifier accepts a ZK proof for the declared circuit on the served transcript:
  \(\mathrm{Verify}(A_{\mathrm{out}}, x, y_{\mathrm{in}}, \pi)=\mathrm{accept}\).
  
  \item \emph{\textbf{S2.} Exact Transcript Alignment.} The served output is exactly reproduced by the declared architecture under ghost weights on the same input and seed:
  \( y_{\mathrm{in}} = f_{A_{\mathrm{out}}}(W_{\mathrm{out}}^{\mathrm{ghost}}, x; s) \).

  \item \emph{\textbf{S3.} Indistinguishability.} For fixed $(A_{\mathrm{out}},x,y)$ and circuit, accepted proofs from hollowed runs are (up to negligible differences) distributed the same way as those from honest runs.

\end{itemize}



\section{Hollow-LLM Attack}
\label{sec:attack}

Building on the proof contract and notation in \secref{sec:framework}, we now instantiate Hollow-LLM attack with two training-free ghost-weight constructions, \emph{Attack~A} and \emph{Attack~B}. Both exploit algebraic invariances of pre-LN Transformer blocks to embed a smaller served \emph{inner} model $A_{\mathrm{in}}$ into a larger, verifiable \emph{outer} architecture $A_{\mathrm{out}}$ while preserving the input–output mapping. 




\subsection{Algebraic Invariances of Transformer Blocks}
\label{sec:invariances}

Naïvely, it is \emph{not} obvious that one can hollow out a Transformer simply by adding layers or replicating coordinates.
Residual connections, LayerNorm, nonlinearities, and multi-head attention tightly couple coordinates, so arbitrary depth or width inflation usually changes the logits (\textit{inner model and committed outer model then have \textbf{different} output}), even if the new parameters look “small” or “unused.”
However, we identified that they nonetheless admit a small set of algebraic invariances that allow rigorously controlled structural manipulation. These invariances characterize exactly when added depth behaves as a structural identity and when width expansion preserves the input–output function. They are the algebraic backbone that makes both Attack A and Attack B possible.
In what follows, we isolate four such invariances. Each appears simple in isolation, but together they determine when a pre-LN Transformer block can be expanded along depth or width without altering its functional behavior. We later use these properties to construct zero-computation layers (Attack A) and width-replicated blocks (Attack B). We adopt the pre-LN architecture from \secref{sec:transformer-arch}.
For a single attention block followed by an FFN block, the pre-LN update is
\[
\begin{aligned}
x_{\ell}' &= x_{\ell}
  + F_{\mathrm{attn},\ell}\bigl(\mathrm{LN}_{\gamma_{\ell}^{\mathrm{attn}},\beta_{\ell}^{\mathrm{attn}}}(x_{\ell})\bigr), \\[4pt]
x_{\ell+1} &= x_{\ell}'
  + F_{\mathrm{ffn},\ell}\bigl(\mathrm{LN}_{\gamma_{\ell}^{\mathrm{ffn}},\beta_{\ell}^{\mathrm{ffn}}}(x_{\ell}')\bigr).
\end{aligned}
\]
Here $x_{\ell}'$ denotes the intermediate representation after the attention sublayer.
The key point is that each sublayer appears only inside a \emph{residual add}.

\bheading{Residual identity.}
Consider a generic pre-LN residual block
\[
  x \;\mapsto\; x + F(\mathrm{LN}(x)).
\]
If we choose parameters so that $F(\mathrm{LN}(x)) = 0$ for all inputs in the support of the deployment (here, the quantized activations that can arise from the inner model under the verifier's arithmetic), then the block acts as an exact identity:
\[
  x \mapsto x + 0 = x.
\]
A stack of such blocks contributes zero effective depth while still being fully compatible with the circuit.
This property allows Attack A to insert arbitrarily many declared layers that contribute symbolic depth but perform no computation.

\bheading{Elementwise nonlinearities.}
Let $\phi$ be any elementwise nonlinearity used in the FFN (e.g., GELU, ReLU, SiLU), realized via lookup tables in the circuit.
For any vector $z$ and replication factor $m\in\mathbb{N}$, define
\[
  R\,z \;\in\; \mathbb{R}^{md}
  \quad\text{by}\quad
  R\,z = (z,\dots,z)
\]
(the concatenation of $m$ identical copies; we call $R$ the \emph{replication operator}).
Then
\[
  \phi(0) = 0,
  \qquad
  \phi(R\,z) = R\,\phi(z)
\]
because $\phi$ is applied coordinatewise.
Thus zeros remain zeros under $\phi$, and replicated coordinates remain exactly equal.
This ensures that padding or replication does not introduce cross-coordinate interference and allows us to zero out padded FFN neurons in Attack~A and to preserve width-wise replication in Attack~B.

\bheading{LayerNorm replicate invariance.}
For a hidden vector $x\in\mathbb{R}^d$, LayerNorm computes
\[
  \mathrm{LN}_{\gamma,\beta}(x)
  =
  \gamma \odot \frac{x-\mu(x)\mathbf{1}}{\sigma(x)} + \beta,
\]
where $\mu(x)$ and $\sigma(x)$ are the per-channel mean and standard deviation, and $\gamma,\beta\in\mathbb{R}^d$ are learned parameters.
Now replicate $x$ into $x' = R\,x\in\mathbb{R}^{md}$ using the same operator $R$ as above.
Since $x'$ consists of $m$ identical length-$d$ blocks, its mean and variance match those of $x$:
\[
  \mu(x') = \mu(x),
  \qquad
  \sigma(x') = \sigma(x).
\]
If we also replicate the LN parameters as
\[
  \gamma' = R\,\gamma,
  \qquad
  \beta'  = R\,\beta,
\]
then LayerNorm commutes with replication:
\[
  \mathrm{LN}_{\gamma',\beta'}(x')
  = 
  R\,\mathrm{LN}_{\gamma,\beta}(x).
\]
This \emph{LayerNorm replicate invariance} is the core algebraic fact that makes Attack~B possible: we can inflate the width by a factor $m$ while preserving the normalized activations that drive attention and FFN computation.

\bheading{Multi-head attention under replication.}
Write a single multi-head attention sublayer as
\[
  \mathrm{MHA}(x)
  =
  W_o \bigl( \mathrm{Concat}(h_1,\dots,h_H) \bigr),
\]
where each head $h_j$ is computed from $x$ via learned projections $W_q^{(j)},W_k^{(j)},W_v^{(j)}$ and a softmax.
For Attack~B, we construct an outer attention with width $d'_{\mathrm{model}} = m\,d_{\mathrm{model}}$ by:
\begin{itemize}
  \item replicating the inner projections across $m$ disjoint subspaces of the wider hidden state, forming a block-diagonal weight matrix, and
  \item replicating the hidden state into these subspaces using $R$.
\end{itemize}

\deleted{
Because the attention equations are linear in the projections and apply per head, each replicated block behaves like a copy of the inner attention.
Combined with LayerNorm replicate invariance, the per-head distributions and outputs are preserved up to exact equality in the circuit’s arithmetic.
, which is formalized when we define Attack~B’s expansion operator.}

\added{Because the attention equations are linear in the projections and apply independently per head, each replicated block behaves as an exact copy of the inner attention. Combined with LayerNorm replicate invariance, this preserves the per-head score distributions and outputs exactly under the circuit arithmetic.

These invariances are precisely the ingredients used below. Attack~A uses residual identity and FFN zero-padding to inflate depth without changing the realized computation. Attack~B uses replication together with block-diagonal attention/FFN structure to inflate width while keeping the logits unchanged.}

\bheading{FFN zero-padding.}
A two-layer FFN with hidden width $d_{\mathrm{ff}}$ computes
\[
  \mathrm{FFN}(x)
  =
  W_2\,\phi(W_1 x + b_1) + b_2.
\]
If we enlarge $d_{\mathrm{ff}}$ by appending extra neurons whose incoming and outgoing weights are zero, then for any input we have
\[
  \phi(W_1' x + b_1')
  =
  \bigl(\phi(W_1 x + b_1),\, 0\bigr),
  \qquad
  W_2' = [W_2 \;\; 0],
\]
so the FFN output remains unchanged.
This lets Attack~A conform to a larger declared FFN width without altering the computation; the added parameters are ghost parameters.

\bheading{From invariances to attacks.}
We will next show that the combination of these invariance is sufficient to instantiate the two Hollow-LLM attacks described below.
In Attack~A (\secref{sec:attackA}), we use the residual-identity property to insert arbitrarily many zero-computation layers that are visible to the verifier but invisible to the serve path.
In Attack~B (\secref{sec:attackB}), we combine replication invariances, LayerNorm behavior, and block-diagonal attention/FFN structure to inflate the model dimension while preserving the logits.


\subsection{Attack A: Hollowing Depth through Zero-Work Residual Blocks}
\label{sec:attackA}

\begin{figure}[t]
  \centering
  \hspace*{0.07\linewidth}%
  \includegraphics[width=0.95\linewidth]{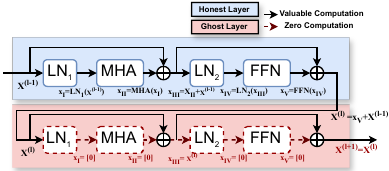}%
  \caption{Attack~A (add-zero-layers). The inner model's layers (solid) are embedded as a prefix of the outer model. The added layers (dashed) are configured so that both the attention and FFN sublayers output zero, leaving the residual path unaffected. Although the verifier observes a depth of $L_{\mathrm{out}}$ and an FFN width of $d_{\mathrm{ff,out}}$, the effective computation still matches that of the depth-$L_{\mathrm{in}}$ inner model.}
  \label{fig:attackA}
\end{figure}

Attack~A increases the declared depth and, optionally, the FFN width of the model while keeping the effective computation depth $L_{\mathrm{in}}$ equal to that of the inner model $A_{\mathrm{in}}$. The served model is embedded as a prefix of the declared layers, and all remaining blocks are configured to behave as algebraic identities. These added layers are fully visible to the verifier but invisible to the serve path (\figref{fig:attackA}).

\bheading{Recipe and subsection structure.}
Formally, we start from an inner architecture $A_{\mathrm{in}}$ with depth $L_{\mathrm{in}}$ and per-layer FFN width $d_{\mathrm{ff,in}}$ and construct an outer architecture $A_{\mathrm{out}}$ with depth $L_{\mathrm{out}}\ge L_{\mathrm{in}}$ and FFN width $d_{\mathrm{ff,out}}\ge d_{\mathrm{ff,in}}$, matching $A_{\mathrm{in}}$ on all other public hyperparameters. \secref{sec:attackA:1} fixes this public shape and defines an expansion operator $E_{\text{layers}}$ that copies the first $L_{\mathrm{in}}$ layers and fills the remaining layers with zero-computation residual blocks. \secref{sec:attackA:2} then shows how a prover who actually runs $A_{\mathrm{in}}$ can edit cached activations into a witness for $A_{\mathrm{out}}$, so that the verifier’s circuit reproduces exactly the inner-model transcript on the same $(x,s)$.

\bheading{Challenge and idea.}
Extra layers usually perturb LayerNorm statistics and residual scaling, altering logits even when their parameters are “small.” Attack~A uses the residual-identity and FFN zero-padding properties from \secref{sec:invariances} to set both the attention and FFN sublayers of added blocks to output exact zeros on the verifier’s arithmetic domain, so each added block acts as $x \mapsto x$ and contributes depth only symbolically.

\subsubsection{Public Shape and Ghost-weight Construction}
\label{sec:attackA:1}
The public outer model architecture describes:
\begin{itemize}
  \item the larger depth $L_{\mathrm{out}}$,
  \item the larger FFN width $d_{\mathrm{ff,out}}$,
  \item identical attention interfaces $(d_{\mathrm{model}}, h, d_k, d_v)$, and
  \item identical embedding and decoder layers $(E,D)$ and positional encoding.
\end{itemize}
Thus the verifier’s circuit is locked to $L_{\mathrm{out}}$ blocks with FFN width $d_{\mathrm{ff,out}}$ and will expect intermediate tensors of that shape in the witness.

\bheading{Expansion operator \(\mathcal{E}_{\mathrm{layers}}\).}
We define a map
\[
  \mathcal{E}_{\mathrm{layers}}
  : W_{\mathrm{in}} \mapsto W_{\mathrm{out}}^{\mathrm{ghost}}
\]
that copies the first $L_{\mathrm{in}}$ layers and fills the remaining layers with identities:
\begin{itemize}
  \item For each layer $\ell \le L_{\mathrm{in}}$, copy the attention and FFN parameters directly,
  \begin{align*}
    (W_q^\ell,W_k^\ell,W_v^\ell,W_o^\ell)^{\mathrm{out}} &= (W_q^\ell,W_k^\ell,W_v^\ell,W_o^\ell)^{\mathrm{in}},\\
    (W_1^\ell,W_2^\ell)^{\mathrm{out}} &= (W_1^\ell,W_2^\ell)^{\mathrm{in}},
  \end{align*}
  with the same biases and LayerNorm parameters.
  If $d_{\mathrm{ff,out}} > d_{\mathrm{ff,in}}$, append rows/columns of zeros and zero biases to match the outer width.
  \item For each added layer $\ell > L_{\mathrm{in}}$, set the attention and FFN sublayers to output zeros for all reachable inputs:
  \begin{align*}
    W_q^\ell &= 0,\quad W_k^\ell = 0,\quad W_v^\ell = 0,\quad W_o^\ell = 0,\\
    W_1^\ell &= 0,\quad b_1^\ell = 0,\quad W_2^\ell = 0,\quad b_2^\ell = 0,
  \end{align*}
  with arbitrary (but fixed) LayerNorm parameters.
\end{itemize}
By residual identity and FFN zero-padding, each added block computes $x\mapsto x$ on the verifier’s arithmetic domain.

\deleted{Therefore the stacked effect of the outer model on inputs of length $\leq L_{\mathrm{in}}$ is exactly that of the inner model, possibly followed by a sequence of identity blocks.}
\added{Therefore, for every supported prompt and decoding seed, the outer model follows exactly the same hidden-state trajectory as the inner model through the first $L_{\mathrm{in}}$ blocks, after which only identity blocks remain. Equivalently,
\[
f_{A_{\mathrm{out}},E_{\mathrm{layers}}(W_{\mathrm{in}})}(x;s)=f_{A_{\mathrm{in}},W_{\mathrm{in}}}(x;s)
\]
for all supported $(x,s)$.}

\subsubsection{Witness Editing (serve → audit)}
\label{sec:attackA:2}

\deleted{At inference serve time, the dishonest provider runs only the inner model (Ain, Win). For each token position t, the provider computes the activations and logits using Lin layers and returns the next-token prediction y.}
\added{At serve time, the dishonest provider runs only $(A_{\mathrm{in}},W_{\mathrm{in}})$ and materializes activations only for the first $L_{\mathrm{in}}$ blocks. For each token position $t$, the provider computes the inner-model activations and logits and returns the next-token prediction $y$.}
Optionally, for anticipated audits the provider caches intermediate activations
\[
  \{x_{\ell,t}\}_{\ell=0}^{L_{\mathrm{in}}}
\]
for each token position in the transcript; these are internal and never exposed to the verifier.

\bheading{Audit path.}
When an audit requires a zero-knowledge proof, the prover must supply a witness consistent with running the declared outer model $(A_{\mathrm{out}},W_{\mathrm{out}}^{\mathrm{ghost}})$.
The witness editing procedure is:
\begin{enumerate}
  \item For layers $\ell \le L_{\mathrm{in}}$, copy all cached intermediate tensors (activations, attention scores, FFN outputs) from the serve-time run into the corresponding slots of the outer-model trace.
  \item For FFN padded neurons (if $d_{\mathrm{ff,out}}>d_{\mathrm{ff,in}}$), set their pre-activations and post-activations to zero, which is consistent with the zero weights and $\phi(0)=0$.
  \item For added layers $\ell > L_{\mathrm{in}}$, set the sublayer outputs to zero and the residual outputs equal to their inputs:
  \[
    x_{\ell,t}^{\mathrm{out}}=x_{\ell-1,t}^{\mathrm{out}},
    F_{\mathrm{attn},\ell}(x_{\ell-1,t}^{\mathrm{out}}) = 0,
    F_{\mathrm{ffn},\ell}(x_{\ell,t}'^{\mathrm{out}}) = 0.
  \]
\end{enumerate}
Because the ghost weights were chosen so that the added blocks are algebraic identities, the verifier’s circuit accepts this witness as a valid execution of $A_{\mathrm{out}}$:
for all $|x|\le L_{\mathrm{in}}$,
\[
  f_{A_{\mathrm{out}},\mathcal{E}_{\mathrm{layers}}(W_{\mathrm{in}})}(x)
  =
  f_{A_{\mathrm{in}},W_{\mathrm{in}}}(x).
\]

\bheading{Takeaway.} \deleted{Attack A increases the depth while leaving the effective computation unchanged; the neutralized layers are visible to the verifier and invisible to the forward pass.}
\added{Attack~A increases the declared depth without changing the realized computation: the extra layers remain part of the verified outer circuit, but they contribute only residual identities during both serving and audit-time witness generation.}


\subsection{Attack B: Width Inflation via Replicated Coordinates}
\label{sec:attackB}

\begin{figure}[t]
  \centering
  \includegraphics[width=\linewidth]{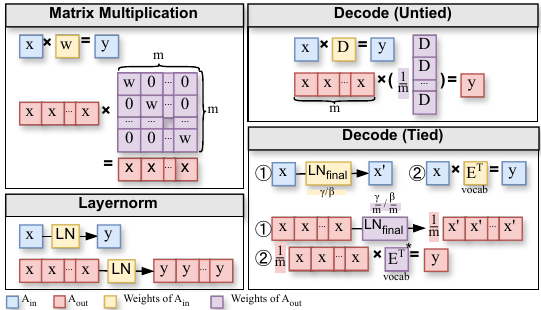}%
  \caption{Attack~B (enlarge-embedding by replication).
  The hidden state is replicated across $m$ blocks of coordinates.
  Attention and FFN weights are arranged in block-diagonal and repeated patterns so that each block behaves like a copy of the inner model.
  The output layer is constructed so that logits match the inner model exactly (zero-padded or averaged decoder for untied, and a final $1/m$ rescaling step for tied).}
  \label{fig:attackB}
\end{figure}

Attack~B mirrors the same template along the width axis: it inflates the declared model dimension by replicating the hidden state across $m$ coordinate blocks, arranges attention and FFN weights in block-diagonal / repeated form to forbid cross-block mixing, and then collapses the replicas at the output so that logits match those of $A_{\mathrm{in}}$ exactly (\figref{fig:attackB}).

\bheading{Recipe and subsection structure.}
Let $A_{\mathrm{in}}$ be a pre-LN Transformer with model dimension $d_{\mathrm{model}}$ and weights $W_{\mathrm{in}}$. We define an outer architecture $A_{\mathrm{out}}$ with expanded width $d'_{\mathrm{model}} = m\,d_{\mathrm{model}}$, the same depth $L$, the same dimension of heads, and the same context-length limit. \secref{sec:attackB:1} introduces the replication operator $R$ and constructs ghost weights $E_{\text{embed}}(W_{\mathrm{in}})$ layer by layer so that all hidden states in the outer circuit take the replicated form $x'_{\ell,t} = R x_{\ell,t}$. \secref{sec:attackB:2} then shows that, given an inner run, simply replicating activations and adding a final $1/m$ rescaling step in the tied-decoder case yields a witness that satisfies every equation in the wider circuit and produces identical logits.

\bheading{Challenge and idea.}
Naïvely widening the model changes dot-product scales and LayerNorm statistics, and tied decoders amplify these changes, so the softmax distribution is no longer preserved. Attack~B combines the LayerNorm replicate invariance and per-block attention/FFN structure from \secref{sec:invariances} with an explicit output collapse (zero-padding or a final $1/m$ rescaling for tied decoders) to keep both logits and softmax outputs equal to those of the inner model.

\subsubsection{Public Shape and Replicated Hidden State}
\label{sec:attackB:1}

At serve-time, the prover runs $A_{\mathrm{in}}$ and maintains hidden states $x_{\ell,t}\in\mathbb{R}^{d_{\mathrm{model}}}$.
At proof-time, the witness must present hidden states $x'_{\ell,t}\in\mathbb{R}^{d'_{\mathrm{model}}}$ to the circuit of $A_{\mathrm{out}}$.
\deleted{We arrange that
\[
  x'_{\ell,t} = R x_{\ell,t}
\quad
\text{for all layers $\ell$ and positions $t\le L_{\mathrm{in}}$},
\]
}
\added{
\text{for every layer $\ell$ and every token position $t$ in the served transcript},
}

i.e., the outer hidden state is always $m$ concatenated copies of the inner hidden state.
The LayerNorm replicate invariance from \secref{sec:invariances} ensures that normalized activations also satisfy this property.

\bheading{Expansion operator \(\mathcal{E}_{\mathrm{embed}}\).}
We now describe how to construct the ghost weights layer by layer.
Let $R=\mathbf{1}_m\otimes I_d$ as before, and let $S_s$ select the $s$-th block of $d$ coordinates from a length-$md$ vector.

\bheading{Embeddings and LayerNorm.}
For the token embedding and positional encoding:
\begin{itemize}
  \item For each token $v$ with inner embedding $E_v\in\mathbb{R}^d$, define
  \[
    E'_v = R E_v \in\mathbb{R}^{md}.
  \]
  \item \deleted{For any learned positional encodings $p_t$, set $p_t' = R p_t$ for all $t\le L_{\mathrm{in}}$.}
  \added{\item For any learned positional encodings $p_t$, set $p'_t = Rp_t$ for every supported token position $t$.}
  \item For each LayerNorm with parameters $(\gamma,\beta)$, set $(\gamma',\beta') = (R\gamma, R\beta)$.
\end{itemize}
By construction and the invariances of \secref{sec:invariances}, all normalized inputs to attention and FFN are replicated.

\bheading{Attention sublayers and softmax.}
For each layer $\ell$ and each head, let $W_q^\ell, W_k^\ell, W_v^\ell$ and $W_o^\ell$ denote the inner projections.
We form the outer projections as block-diagonal replications:
\begin{align*}
  W_q'^{\,\ell} &= \mathrm{blkdiag}(W_q^\ell,\dots,W_q^\ell),\\
  W_k'^{\,\ell} &= \mathrm{blkdiag}(W_k^\ell,\dots,W_k^\ell),\\
  W_v'^{\,\ell} &= \mathrm{blkdiag}(W_v^\ell,\dots,W_v^\ell).
\end{align*}
Given a replicated input $x' = R x$, each block $S_s x'$ computes exactly the same queries, keys, and values as the inner head:
\[
  Q_s = W_q^\ell x,
  \quad
  K_s = W_k^\ell x,
  \quad
  V_s = W_v^\ell x
  \qquad\text{for all } s\in[m].
\]

Crucially, the softmax in multi-head attention is taken over the \emph{key positions}, not over the feature dimension.
Replication only happens along the feature dimension: we do not create extra tokens or extra time steps.
Therefore, for each head and each position, the score matrices
\[
  \frac{Q_s K_s^\top}{\sqrt{d_k}}
\]
and the resulting attention weights $\mathrm{softmax}(\cdot)$ are identical across all $s$ and coincide with the inner model’s attention weights.
The concatenation of heads simply glues these replicated outputs along the feature axis.

For the output projection, we can either:
\begin{itemize}
  \item \emph{single-block routing:}
  \[
    W_o'^{\,\ell} = W_o^\ell S_1^\top,
  \]
  i.e., we apply $W_o^\ell$ to the first replicated block and ignore the others; or
  \item \emph{averaging across blocks:}
  \[
    W_o'^{\,\ell}
    =
    \frac{1}{m}\bigl[W_o^\ell \;\; W_o^\ell \;\; \cdots \;\; W_o^\ell\bigr],
  \]
  which averages the $m$ copies of the same head output.
\end{itemize}
In both cases, the attention sublayer output matches the inner attention output exactly for replicated inputs, and the softmax distributions over positions are unchanged.

\bheading{FFN sublayers.}
For each FFN layer, we replicate both the input and hidden dimensions:
\begin{align*}
  W_1'^{\,\ell} &= \mathrm{blkdiag}(W_1^\ell,\dots,W_1^\ell),\\
  W_2'^{\,\ell} &= \begin{bmatrix} W_2^\ell & W_2^\ell & \dots & W_2^\ell \end{bmatrix}.
\end{align*}
Given a replicated input $x' = R x$, we have
\[
  W_1'^{\,\ell} x'
  =
  R (W_1^\ell x),
  \qquad
  \phi(W_1'^{\,\ell}x' + b_1')
  =
  R \phi(W_1^\ell x + b_1),
\]
and then
\[
  W_2'^{\,\ell}\,\phi(W_1'^{\,\ell}x' + b_1')
  =
  W_2^\ell\,\phi(W_1^\ell x + b_1),
\]
so the FFN output exactly matches the inner FFN.
Again, zeros and replication are preserved by the elementwise nonlinearity.

\bheading{Output logits and softmax.}
Finally, we must ensure that the \emph{logits} and therefore the softmax outputs seen by the verifier match those produced by the inner model.
This is the place where naive “just multiply everything by $m$” reasoning breaks: changing the scale of logits will in general change the softmax distribution, and the circuit checks those exact arithmetic relations.
We must therefore design $A_{\mathrm{out}}$ so that its logits are \emph{exactly} the same as the inner logits on all supported transcripts.

We consider untied and tied decoders separately.

\medskip
\noindent
\textbf{Untied decoder.}
If the output projection $D$ is not tied to $E$, we have full freedom to choose $D'$ without affecting embeddings.
We maintain the replicated hidden state $h' = R h$ and define $D'$ so that $D'h' = D h$ coordinatewise.

Two simple constructions satisfy this:

\begin{itemize}
  \item \emph{Zero-padded decoder:}
  \[
    D' = \bigl[ D \;\; 0 \;\; 0 \;\; \cdots \;\; 0 \bigr],
  \]
  i.e., $D$ applied to the first block and zeros on the remaining $(m-1)$ blocks.
  Then
  \[
    D' h'
    =
    D h + 0 + \cdots + 0
    =
    D h.
  \]

  \item \emph{Averaging decoder:}
  \[
    D'
    =
    \frac{1}{m}\bigl[ D \;\; D \;\; \cdots \;\; D \bigr].
  \]
  Since each block of $h'$ is equal to $h$, we obtain
  \[
    D' h'
    =
    \frac{1}{m}\sum_{s=1}^m D h
    =
    D h.
  \]
\end{itemize}

In both cases, the logits supplied to the softmax are \emph{identical} to the inner logits.
Therefore, the softmax outputs and the next-token distribution are unchanged, and the verifier sees exactly the same equations as in an honest run of $A_{\mathrm{in}}$.

\medskip
\noindent
\textbf{Tied Decoder.}
When $D = E^\top$, the situation is more constrained.
We still set $E'_v = R E_v$ so that embeddings respect replication, and we tie $D' = E'^\top$ as usual.
For a replicated hidden state $h' = R h$, the raw dot product logits satisfy
\[
  \langle E'_v, h' \rangle
  =
  \langle R E_v, R h \rangle
  =
  m\,\langle E_v, h\rangle,
\]
so naive tying introduces a uniform factor $m$ on all logits.
That factor \emph{does} change the softmax (it sharpens the distribution) and breaks equality at the level of probabilities.

To fix this while preserving tying, we rescale the final hidden representation by $1/m$ \emph{right before} applying the tied decoder.
Concretely, let $z_L$ be the pre-readout hidden state of the inner model and $z'_L = R z_L$ its replicated counterpart.
In the outer model we apply an extra scalar map $\alpha I$ with $\alpha = 1/m$ to the final LayerNorm output:
\[
  \tilde{z}'_L = \alpha\, z'_L = \frac{1}{m} R z_L.
\]
We can fold this $\alpha$ into the scale parameter of the \emph{last} LayerNorm or implement it as an explicit diagonal linear layer that is part of $A_{\mathrm{out}}$.
The logits then satisfy
\[
  \langle E'_v, \tilde{z}'_L \rangle
  =
  \Big\langle R E_v, \frac{1}{m} R z_L \Big\rangle
  =
  \frac{1}{m} \, m \, \langle E_v, z_L \rangle
  =
  \langle E_v, z_L \rangle,
\]
so tying is preserved and the logits are exactly equal to those of the inner model.

It is important that this $1/m$ rescaling happens at the \emph{last} LayerNorm (or immediately after the final residual block).
If we attempted to scale at the very beginning of the network, pre-LN residual structure
\[
  x_{\ell+1} = x_\ell + F_\ell(\mathrm{LN}(x_\ell))
\]
would reintroduce unscaled contributions through the residual adds: different paths would carry differently scaled signals, and their sums would no longer match the inner representation.
Only after the final residual add—when no further mixing occurs—can we safely apply a uniform scalar factor and preserve equality with the inner model.

\subsubsection{Witness Editing and Correctness}
\label{sec:attackB:2}
Similar as Attack~A, the prover actually runs $A_{\mathrm{in}}$ and caches the inner activations.
To construct the witness, the prover:
\begin{enumerate}
  \item replicates each inner hidden state $x_{\ell,t}$ into $x'_{\ell,t} = R x_{\ell,t}$;
  \item uses the ghost weights from $\mathcal{E}_{\mathrm{embed}}(W_{\mathrm{in}})$ for all sublayers; and
  \item fills in the attention and FFN intermediates implied by the replicated computation, including the final $1/m$ rescaling step in the tied-decoder case.
\end{enumerate}
By the invariances already established and the explicit treatment of the logits above, every layer equation in the outer circuit is satisfied, and both logits and softmax outputs match those of the inner model exactly.

\begin{figure}[t]
  \centering
  \includegraphics[width=0.99\linewidth]{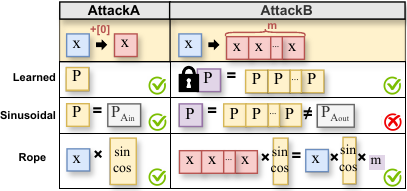}%
  \caption{Positional encodings under both attacks.
  For learned encodings, both attacks can reuse or replicate the same table.
  For deterministic sinusoidal encodings, Attack~A still matches the inner model, but Attack~B cannot: the circuit recomputes $P_{\text{out}}$ as a public function of $(t,d_{\text{model,out}})$, so in general $P_{\text{out}}\neq R P_{\text{in}}$ and the mismatch is ZK-detectable.
  For RoPE, we align rotation frequencies and replicate heads so that within-head indices match and rotations coincide.}
  \label{fig:posenc}
\end{figure}

\bheading{Takeaway.} Attack~B increases the \emph{reported} width by replicating coordinates and then collapsing them at the output, so the verifier sees a larger model but the logits and serve‑path cost match $A_{\text{in}}$.


\subsection{Positional Encodings}
\label{sec:pe}

Unlike most other components of the Transformer, whose basic blocks have remained fairly stable, positional encoding has evolved substantially over time. Early models used fixed sinusoidal functions; later variants moved to learned absolute tables; and modern LLMs increasingly adopt rotary encodings (RoPE) and related schemes. In this subsection we examine these major choices and analyze, for each, how compatible it is with our hollow LLM constructions.

Because the positional scheme is hard-coded in the verifier’s circuit, both attacks must respect whatever encoding the outer architecture specifies. \figref{fig:posenc} summarizes the three cases: learned encodings and RoPE remain compatible with both attacks after simple copying or replication, whereas deterministic sinusoidal encodings still support Attack~A but can rule out Attack~B.

\bheading{Scope and Placement.}
We assume $|x|\le L_{\mathrm{in}}$, consistent with the inner model’s context length.
Under pre-LN, positional information is added before the first LayerNorm in each layer, so Attack~A’s zero blocks and Attack~B’s replication always act on states that already include positions.
We only need the embedding + position step to respect the invariances in \secref{sec:invariances}.

\bheading{Learned absolute tables.}
For learned absolute embeddings $p_t\in\mathbb{R}^{d_{\mathrm{model}}}$:
\begin{itemize}
  \item \emph{Attack~A.} The outer model reuses the same table for $t\le L_{\mathrm{in}}$; added layers do not change $p_t$.
  \item \emph{Attack~B.} We define $p_t' = R p_t$, so that $x_0' = R(E(x_t)+p_t)$ and the replicate-invariance conditions hold.
\end{itemize}

\bheading{Sinusoidal encodings.}
For deterministic sinusoidal encodings, the circuit computes $p_t = g(t,d_{\mathrm{model}})$ as a fixed public function.
Once $d_{\mathrm{model,out}}$ is fixed, $P_{\text{out}}$ is fully determined.
\begin{itemize}
  \item \emph{Attack~A.} Since $d_{\mathrm{model,out}} = d_{\mathrm{model,in}}$, we have $p_t^{\text{out}} = p_t^{\text{in}}$ for $t\le L_{\mathrm{in}}$, and extra layers leave positions unchanged.
  \item \emph{Attack~B.} With $d_{\mathrm{model,out}} = m\,d_{\mathrm{model,in}}$ and $m>1$, the circuit enforces $p_t' = g(t,d_{\mathrm{model,out}})$, while we would need $p_t' = R\,g(t,d_{\mathrm{model,in}})$ for the replicate construction.
  In general $g(t,d_{\mathrm{model,out}})\neq R\,g(t,d_{\mathrm{model,in}})$, so the witness cannot satisfy both the circuit equation and the desired replicated hidden state.
  The resulting mismatch at the first layer makes Attack~B ZK-detectable (red cross in \figref{fig:posenc}). 
\end{itemize}

\bheading{RoPE (rotary encodings).}
RoPE rotates queries/keys according to frequency-specific angles.
To preserve RoPE under Attack~B we:
\begin{itemize}
  \item use the same rotation frequencies in each replicated head; and
  \item apply RoPE identically in each replicated block of coordinates.
\end{itemize}
If the inner head uses rotation matrix $R_t$ at position $t$, the outer model applies $R_t$ independently in each block.
Attention scores and probabilities are then identical across blocks and equal to those of $A_{\mathrm{in}}$.

\bheading{Takeaway.}
Learned tables and RoPE-style schemes can be aligned with both attacks by copying or replicating positional vectors in a way that respects \secref{sec:invariances}.
Deterministic sinusoidal encodings, when fixed as a public function of $(t,d_{\mathrm{model}})$, block the widen-by-replication trick of Attack~B, but leave Attack~A and the learned/RoPE variants of Attack~B intact.
\added{This is therefore a limitation of one particular construction rather than of Hollow-LLM more broadly. Fixed sinusoidal encodings prevent our width-expansion route, but they do not affect Attack~A, and Attack~B remains available under learned or RoPE-style positional schemes. Since many modern LLMs no longer use fixed sinusoidal encodings, this restriction narrows one variant without substantially reducing the practical relevance of the attack.}

\section{Evaluations}
\label{sec:experiments}


Beyond theoretical analysis, our evaluation aims to empirically corroborate the core claim of Hollow-LLM attack.
Our experiments pursue three questions:
(1) whether serve-side metrics (compute cost, latency, and memory cost) remain invariant
when $A_{\mathrm{in}}$ is fixed, even as $A_{\mathrm{out}}$ changes;
(2) whether zero-knowledge proving costs scale with the size of the declared
model rather than the executed one; and
(3) whether each structural transformation of attacks modifies only the symbolic declared circuit while leaving the executed computation invariant, and whether these transformations compose cleanly without introducing cross-effects.

\added{Our evaluation focuses on serve/prove asymmetry and exact circuit preservation; deployment-level detectability is discussed separately in \secref{sec:discussion}.}.

\bheading{Evaluation Methodology.}
To answer these questions, we evaluate each configuration end-to-end: serving a request, generating a proof, and verifying the proof. For every declared architecture, we measure serve-side behavior, proving cost, and verification correctness while keeping the executed inner model fixed. By isolating the contributions of $A_{\mathrm{in}}$, $A_{\mathrm{out}}$, and their structural differences, the methodology directly tests the causal dependencies of our design.

\subsection{Experimental Setup}
\label{sec:setups}

\bheading{Evaluation Model.} 
Our design involves two models: (1) the \emph{executed} inner model  ($A_{\mathrm{in}}$) during serving, which determines the computation used to serve a request, and
(2) the \emph{declared} outer model ($A_{\mathrm{out}}$), which determines the
public circuit used for zero-knowledge proving.  Our evaluation makes this
separation explicit by fixing $A_{\mathrm{in}}$ while systematically varying
$A_{\mathrm{out}}$.
This serve/prove separation enabled by the Hollow-LLM attack is agnostic to the underlying model family and is expected to hold across a wide range of scale ratios (guidelines for choosing a ratio are discussed in~\secref{sec:discussion}).
For evaluation, we adopt the ZKP procedure from zkGPT~\cite{qu:2025:zkgpt} as a concrete ZKP-verified LLM inference setting. Following the zkGPT framework, the prover executes a compact 6-layer, 512-dimensional Transformer, which we denote $IM_1$ as our baseline inner model.  
All serving-side costs, including FLOPs, memory traffic, and latency, are fully determined by this executed model. 
In contrast, the declared outer architecture is inflated through structural transformations that alter its symbolic shape while preserving its functional behavior. 
For this inner model, we instantiate a single representative scale ratio for Attack~A, Attack~B, and their composition A+B, introduced in~\secref{sec:attack}:
\begin{itemize}
    \item \textbf{Attack~A (add-zero-layers):}
    increases the declared depth by inserting identity residual blocks
    (6 layers $\rightarrow$ 12 layers).  
    The executed inner model, denoted $IM_1$, remains unchanged.

    \item \textbf{Attack~B (enlarge-embedding):}
    widens the declared hidden and embedding dimensions
    (512 $\rightarrow$ 1024) while still executing the same $IM_1$ model.

    \item \textbf{A+B composition:}
    applies both depth and width expansion, yielding a declared architecture
    (12 layers, 1024 hidden) structurally equivalent to a larger transformer.
    Execution remains fixed at $IM_1$; only the declared circuit grows.
\end{itemize}

This fixed-$A_{\mathrm{in}}$, varied-$A_{\mathrm{out}}$ design allows us to test the claim: serve-side metrics should track only the executed computation, while proving costs should scale only with the declared circuit.






\begin{table*}[t]
\centering
\small
\caption{\textbf{Serve-side and ZKP cost under depth/width expansions.}
Serve-side time is normalized to the baseline inner model $IM_1$ (6 layers, 512-dim).  Declared parameters reflect the public circuit after applying Attack~A/B/A+B.  ZKP costs scale only with the declared architecture, as the executed model is fixed.}
\label{tab:full-scaling}
\begin{tabular}{llccccc cccc cc}
\toprule
& & 
\multicolumn{2}{c}{\textbf{Declared Params}} &
\multicolumn{2}{c}{\textbf{Serve Time}} &
\multicolumn{4}{c}{\textbf{ZKP Costs}} &
\multicolumn{2}{c}{\textbf{Theoretical FLOPs}} \\
\cmidrule(lr){3-4}\cmidrule(lr){5-6}\cmidrule(lr){7-10}\cmidrule(lr){11-12}
\textbf{Inner} & \textbf{Attack} &
\textbf{Layers} & \textbf{Embed} &
\textbf{Prefill} & \textbf{Decode} &
\textbf{Gates} & \textbf{Prover} & \textbf{Verify} & \textbf{Proof KB} &
\textbf{Prefill} & \textbf{Decode} \\
\midrule
$IM_1$ & --     & 6  & 512  & 1$\times$ & 1$\times$ & 
1$\times$   & 1$\times$ & 1$\times$ & 45.6 &
1$\times$ & 1$\times$ \\

$IM_1$ & A      & 12 & 512  & 1$\times$ & 1$\times$ & 
1.99$\times$ & 1.69$\times$ & 1.72$\times$ & 83.5 &
1.4$\times$ & 2.0$\times$ \\

$IM_1$ & B      & 6  & 1024 & 1$\times$ & 1$\times$ &
2.95$\times$ & 1.87$\times$ & 1.30$\times$ & 48.2 &
2.8$\times$ & 3.9$\times$ \\

$IM_1$ & A+B    & 12 & 1024 & 1$\times$ & 1$\times$ &
5.89$\times$ & 4.81$\times$ & 2.48$\times$ & 88.3 &
4.5$\times$ & 7.8$\times$ \\

$IM_2$ & --     & 12 & 1024 & 2.4$\times$ & 3.1$\times$ &
5.89$\times$ & 4.74$\times$ & 2.56$\times$ & 88.3 &
4.5$\times$ & 7.8$\times$ \\
\bottomrule
\end{tabular}
\end{table*}

\bheading{Workloads and Testbed.} 
We use fixed-length prompts of $T=64$ tokens and greedy decoding.  This choice
intentionally isolates the structural effects of declared-model transformations:
both Attack~A and Attack~B modify only the shape of the public circuit and are
insensitive to prompt semantics, token distribution, or linguistic variation.
Thus, short deterministic prompts are fully sufficient to exercise the
serve/prove split that attacks aim to validate.
We further use a compact transformer as the executed model $A_{\mathrm{in}}$.
This choice is deliberate: the purpose of our evaluation is to verify the
\emph{correctness and structural separation} of serving and proving, not to
stress-test large-model throughput.  Since transformations operate only
on $A_{\mathrm{out}}$ and not on the runtime semantics of $A_{\mathrm{in}}$,
evaluating correctness does not require a large or instruction-heavy model.
Using a small model enables complete end-to-end proving while keeping the
methodology controlled and fully reproducible. All experiments are conducted on a local workstation running Ubuntu 24.04 with Intel Core Ultra 7 265F CPU (20 cores).

\subsection{Results}
\label{sec:results}

Our experiments vary the declared depth and width while keeping the executed inner model fixed, allowing us to isolate three properties of interest: (1) whether serve-side latency depends only on the executed model, (2) whether zero-knowledge proving cost scales with the declared architecture, and (3) whether structural manipulations exhibit locality and predictable composition.

Each row in Table~\ref{tab:full-scaling} represents a pair of executed and declared architectures.  The inner column denotes the model actually executed during serving (e.g., $IM_1$ is a 6-layer, 512-dimension GPT-2 backbone).  The outer column specifies the declared public architecture obtained by applying Attack~A (depth expansion), Attack~B (width expansion), or their composition.  These structural changes modify only the shape of the declared circuit; the executed inner model remains fixed for all rows.
Serve-side columns report latency relative to the baseline.  ZKP Costs columns report the ratio of gate size, prover time, and verification time induced by the declared model.  
Thus, serve-side ratios reflect properties of the executed model, whereas ZKP ratios reflect properties of the declared circuit.  This separation makes it easy to read how each attack influences serving and
proving, as discussed below.

\bheading{Serve-side Invariance.}
Serve-side behavior is identical across all configurations because the attack always executes the same inner model.  
This is visible in Table~\ref{tab:full-scaling}: the Prefill and Decode columns remain at 1x for all A, B, and A+B rows, even though their declared models differ in depth and width.  
Only the honest large model shows higher serve-side ratios, confirming that serve-side cost depends solely on the executed model and is unaffected by changes to the declared architecture.

\bheading{Declared-model Scaling.}
Zero-knowledge cost scales with the size of the declared architecture: increasing the declared depth or width enlarges the public circuit and raises gate count, proving time, and verification cost.  
In Table~\ref{tab:full-scaling}, Attack~A and Attack~B each increase different parts of the ZKP cost, where A adds depth-related overhead and B contributes width-related overhead. Their effects differ but are both strictly above the baseline.  
The A+B configuration combines both forms of expansion and therefore shows the largest ZK cost, even though the executed model remains fixed.  
Since the inner model never changes, all variation originates solely from the declared architecture.

\bheading{Locality and Composition.}
Depth and width expansions influence separate parts of the declared circuit, and their zero-knowledge overheads accumulate predictably.  
In Table~\ref{tab:full-scaling}, Attack A increases only the Gates and proving cost associated with additional layers, while Attack B increases only the costs associated with wider embeddings.  
The A+B row combines both increases, matching the sum of the A and B effects and showing no change in serve-side columns, confirming that these expansions act locally and compose without interference.

\bheading{Comparison to Honest Large Model.}
The full table shows that the A+B configuration reproduces the proving cost of
the honest large reference model ($IM_2$, a 12-layer, 1024-dimensional Transformer).  
The A+B and $IM_2$ rows exhibit nearly identical gate counts, prover time, and
verification time, indicating that the declared circuit obtained from A+B
matches the structural complexity of a large model, even though the system
continues to execute only the compact inner model $IM_1$. 

This creates a compute asymmetry visible directly in the serve-side columns:
$IM_2$ incurs substantially higher prefill and decode cost, while A+B remains at
$1\times$ because it still executes $IM_1$.  
Thus, the attack exposes large-model proving complexity while avoiding the
execution overhead of a large model.

\bheading{Semantic Squivalence.}
All variants preserve the behavior of the executed model. Declared-model expansions introduce only structural changes to the public circuit and do not modify the inner model's forward computation, yielding identical outputs across all configurations.  We confirmed this by observing unchanged perplexity across all experiments.

\bheading{Summary of Findings.}
Together, these results demonstrate the attack's serve/prove separation: serve-side cost is fixed by the inner model, proving cost scales with the declared model, and structural expansions compose cleanly.  Hollow-LLM attack can thus expose large-model proving complexity while executing only a compact model.

\section{Discussion}
\label{sec:discussion}
\subsection{Scope of the Attack}

\bheading{ZKP is Correct but an Overlooked Effort Gap.}
It is important to clarify that the Hollow-LLM attack does not imply any flaw in zero-knowledge proof technology, which remains fundamentally sound as a cryptographic primitive. In ZK LLM serve contexts, the proof mechanism correctly verifies that a given output is consistent with the computations defined by the declared model architecture. However, ZKPs do not attest to the actual effort or computational cost expended to produce that output during the procedure.
The Hollow-LLM attack exploits this distinction: an adversary can supply a verifiably correct result (satisfying all equation-level checks) without performing the expected full-scale computation. This highlights the distinction between ensuring output correctness and ensuring execution effort. Any apparent overpromising stems not from the ZKP technology itself, but from deployment-layer assumptions. \deleted{The mismatch arises from how proofs are interpreted in practice, rather than any failure of the proof system.}
\added{In particular, the mismatch arises when that commitment-relative guarantee is interpreted as evidence of commensurate execution effort, rather than from any failure of the proof system.}

\bheading{Choosing the Appropriate Outer Model Size.} In principle, our ghost weight method allows the declared outer model size to be inflated arbitrarily without changing its input–output behavior. 
However, an overly aggressive enlargement can undermine stealth. A declared architecture that vastly exceeds what users expect for a given service or device may break immersion and raise suspicion. 
\deleted{The attacker must balance the cost-saving incentive of a larger outer model against the risk of detection through anomalous performance or context mismatch.}
\added{The attacker must therefore balance the cost-saving incentive of a larger outer model against the risk of detection through anomalous quality, performance, or deployment mismatch.}
At the same time, users cannot directly observe the true model size or workload, even benchmarking is not a reliable estimator of parameter count~\cite{gao2024model,cai2025you,hinton2015distilling, sanh2019distilbert}, choosing the outer model size thus becomes a domain-specific perception-management problem. The declared outer model should remain plausible given the deployment context. For example, a billions-parameter flagship model might be credible in a cloud API from a major provider, whereas claiming the same size on a lightweight edge device would defy expectations. 
\added{We do not claim that arbitrary over-claiming is indistinguishable. Extreme outer/inner gaps may create detectable quality shifts or behavioral anomalies. The practically relevant regime is modest over-claiming, where even \added{relatively small} savings can be economically meaningful at serving scale while producing weaker external evidence for detection.} 

Determining the exact threshold where enlargement breaches user trust is beyond the scope of the paper, as we focus on demonstrating that enlarged models with equivalent observable behavior can be constructed. Future work or operational policy must address how large an expansion can go undetected in various real-world scenarios.

\bheading{Distinguishing Hollow-LLM Attack from Optimization Techniques.}  Optimization techniques such as KV caching reduce inference cost by reusing computations. For example, using a KV cache skips redundant recomputation of earlier tokens’ effects. However, in benign optimization scenarios, the model preserves its advertised structure and behavior and is simply executed more efficiently. The Hollow-LLM attack, on the other hand, fundamentally violates this fidelity. Instead of accelerating a legitimate large model, the provider actually runs a different, smaller model while presenting it as a larger one. This means the system is no longer faithfully executing the advertised workload; it only mimics the outputs of the large model. Crucially, ZK proof-based verification  will still pass because it checks only that the outputs are consistent with some execution of the declared architecture and does not verify that the full computational effort implied by that architecture was actually expended.


\subsection{Countermeasures}
One potential direction for mitigating the Hollow-LLM attack is to use zero-knowledge proofs not only to verify inference correctness but also to certify structural properties of the model, such as the sparsity or activity of its weight matrices. For example, a prover might demonstrate that certain matrices are not entirely zero. However, this approach faces two major limitations. First, adding sparsity-related subproofs for each parameter matrix introduces substantial overhead and undermines system scalability. Second, merely proving that weights are non-zero is insufficient to rule out Hollow-LLM constructions. An attacker may still be able to design ghost weights that satisfy non-triviality constraints while preserving the same low-effort computation path. A proof that no alternative ghost-weight constructions exist would be required, yet such a guarantee may be generally difficult to achieve.

A practical near-term strategy is to raise the adversary’s cost, but it still cannot offer a fundamental security guarantee.
For example, auditors can use behavioral techniques to raise the cost of cheating and increase detectability. These include challenge-based audits that introduce complex prompts only during verification, tests that temporarily disable attention heads or feedforward channels to measure ablation resilience, and diversity probes that evaluate output variability across decoding seeds or paraphrased prompts. Collectively, these techniques could stress-test the model's response surface and make low-effort deployments more risky.
Another line of defense is to rely on hardware-based mechanisms such as Trusted Execution Environments (TEEs) instead of purely verification. For example, model training or inference can be executed inside a trusted hardware enclave, with hardware isolation and remote attestation providing stronger runtime monitoring and guarantees. Notably, with the recent trend that server-grade GPUs also support emerging GPU-TEE features~\cite{nvidia:2023:h100}, a combined CPU–GPU TEE pipeline can enforce that both the embedding flow and tensor-core computations are executed within measured, attested hardware. 
Nonetheless, TEEs still inherit limitations: they lack public verifiability, depend on proprietary hardware, and remain vulnerable to side channels or firmware-level exploits. These constraints ultimately restrict their applicability in open, decentralized, or multi-party verification settings.

\section{Related Work}

\deleted{\bheading{Towards Verifiable LLM.}}
\added{\bheading{Verifiable LLM Inference.}}
Recent zero-knowledge machine learning (zkML) systems such as zkLLM~\cite{sun:2024:zkllm}, zkGPT~\cite{qu:2025:zkgpt}, and ZkTorch~\cite{chen2025zktorch} demonstrate the feasibility of verifying LLM inference via succinct proofs, even for models with tens of billions of parameters. 
For example, zkLLM proves a 13B-parameter model in under 15 minutes, and zkGPT verifies GPT-2 inference in roughly 25 seconds. General-purpose compilers such as ZKML and ZKTorch further extend verifiable inference to larger models through optimized circuits and proof aggregation. This trustless verification is especially advantageous in distributed or decentralized AI services, where third parties can independently validate the correctness of the model’s outputs. \added{Complementary model-oversight work such as WAVE~\cite{xu2026wave} instead uses hardware-level observations to check whether observed execution is consistent with the claimed model, targeting deployment-level detection rather than the proof guarantee itself.}

Alternatively, trusted execution environments (TEEs) are a popular and industry-ready solution. TEEs isolate model inference within secure hardware enclaves, enabling near-native latency because computation runs on dedicated hardware with minimal overhead. Earlier TEE-based verifiable AI frameworks are usually based on a CPU TEE~\cite{tramer:2018:slalom,hashemi:2021:darknight}, such as Intel SGX. For example, Slalom~\cite{tramer:2018:slalom} partitions neural network layers between an SGX enclave and a GPU, achieving 6× to 20× speedups for verifiable inference compared to enclave-only execution. More recently, NVIDIA released confidential GPU support~\cite{nvidia:2023:h100}, which provides a GPU TEE capability on the H100, B200, and RTX Pro 6000. GPU TEEs can confidentially execute LLM inference with modest performance overhead by encrypting data in use. These TEE-based methods require no code changes and provide real-time protected execution.
However, unlike ZK proofs that offer cryptographic guarantees of correct execution, TEEs rely on specialized hardware and trust assumptions and do not provide public verifiability. Moreover, recent research show that TEE are inherently hardware-dependent and may be vulnerable to microarchitectural attacks~\cite{van:2024:sok,moghimi:2023:downfall,van:2020:sgaxe,van:2023:sgx,chen:2019:sgxpectre,wang:2017:leaky}. 
In scenarios that require strong trustless guarantees or open verification, ZK-proof–based approaches provide a more robust, although higher-latency, alternative.

\added{\bheading{Model-substitution detection and statistical auditing.}
Another related direction asks whether a remote service is actually serving the claimed model by comparing its behavior against external references or statistical fingerprints. Prior work such as VeriLLM, DetectLLM, SVIP ~\cite{wang2025verillm,su2023detectllm,sun2024svip} studies model integrity through output distributions, hidden-state or logit sampling, rank-based tests, equality-style checks, or side observations such as timing and traffic patterns. These methods are especially natural when the target model, its weights, or an official reference implementation are publicly available, because the auditor can compare the deployed service against known ground truth. Our setting is different. Hollow-LLM studies proprietary-weight zkLLM deployments in which the verifier sees only the public architecture, a commitment to private weights, and a proof transcript. In this commitment-only regime, behavioral auditing is best viewed as a complementary defense rather than part of the proof guarantee itself. Accordingly, our contribution is not to replace statistical verification, but to show that standard ZK verification alone does not rule out low-effort witnesses even when every proof remains valid.}

\added{\bheading{Weight manipulation and inactive-layer constructions.} Prior work has shown that carefully structured parameters can make part of a model's nominal computation ineffective. A particularly relevant example is~[46], which studies federated learning and shows how model inconsistency can suppress client-side updates, weakening the intended privacy protection of secure aggregation. Our work follows a related intuition in the setting of zero-knowledge LLM inference: simple zeroed and structured ghost weights can preserve proof validity for a larger declared model while allowing the provider to carry out only the serve-time computation of a much smaller inner model. In this sense, our contribution is to make the same broad phenomenon concrete for zkLLM serving, where the central gap is not update privacy but the lack of effort binding between a valid proof and the computation actually performed.}
\section{Conclusion}
In this paper, we identified and formalized an effort gap in ZK-verified LLM inference, showing that existing systems certify equation-level correctness for a declared architecture but remain agnostic to the computational effort actually expended. We instantiated this gap through the Hollow-LLM attack, in which a dishonest provider masks a smaller inner model behind ghost weights that preserve the public parameter count and zero-knowledge proofs while routing almost all computation through the cheaper model. We further developed concrete, training-free algebraic constructions that exploit transformer invariances to decouple advertised depth and width from true per-token cost, and we analyzed their security and economic impact in realistic deployment settings. Our findings underscore that proof validity alone is insufficient to guarantee faithful execution effort and challenge emerging narratives around zkML-based assurance. 

\added{\section*{Acknowledgment}}
\added{The authors would like to thank Professor Jiapeng Zhang and Xinyu Mao for their insightful discussions and valuable comments. We also thank Haoxuan Xu for his feedback and support throughout the development of this project. Finally, we thank the anonymous reviewers and our shepherd for their constructive guidance during the revision process.}

\bibliographystyle{IEEEtran}
\bibliography{references} 

\section*{Ethics considerations}
Our work is a methodological and systems-oriented study of zero-knowledge verified LLM inference. We formalize the Hollow-LLM attack, construct algebraic ghost-weight families, and evaluate their impact using self-hosted transformer models and zkML pipelines in a controlled research setting; we do not involve human subjects, user studies, or any form of personal or sensitive data.
The main ethical concern is that our constructions could be abused by a dishonest provider in practice to under-provision compute while still passing zero-knowledge verification. However, to the best of our knowledge, no company offers commercial real-time ZKP LLM inference verification service. Making the attacks explicit before real-world deployments emerge is intended to help standardize defenses and evaluation practices, not to enable threats. 

\addedd{\section*{Conflict of Interest Statement}}
\addedd{The authors declare that they have no competing financial or non-financial interests related to this work.}

\section*{LLM usage considerations}
\begin{itemize}
  \item \textbf{Originality.}
  LLMs were used for editorial purposes in this manuscript, and all outputs were inspected by the authors to ensure accuracy and originality. All core technical contributions (formal definitions, attack constructions, proofs, threat model, experiments) and the literature review were developed and written by the authors.

  \item \textbf{Transparency.}
  LLMs are not part of the paper's methodology: they were not used to design attacks, implement systems, or generate or analyze results. Reproducibility of our findings does not depend on access to any particular LLM service.

  \item \textbf{Responsibility.}
  LLM usage was limited to light text editing and did not involve training new models, collecting additional data, or processing sensitive or proprietary content.
\end{itemize}

\newpage 


\appendices

\added{\section{Meta-Review}}

\added{The following meta-review was prepared by the program committee for the 2026
IEEE Symposium on Security and Privacy (S\&P) as part of the review process as
detailed in the call for papers.}

\added{\subsection*{Summary}}

\added{This paper identifies an ``effort gap'' in current Zero-Knowledge Machine
Learning (zkML) protocols, where proofs verify mathematical correctness but not
the actual computational work performed. The authors introduce the
``Hollow-LLM Attack,'' using elegant algebraic constructions called ``ghost
weights'' to allow a provider to serve a small model while providing valid
zero-knowledge proofs for a much larger one.}

\added{\subsection*{Scientific Contributions}}
\begin{itemize}
    \item \added{Identifies an Impactful Vulnerability.}
    \item \added{Provides a Valuable Step Forward in an Established Field.}
\end{itemize}

\added{\subsection*{Reasons for Acceptance}}
\begin{enumerate}
    \item \added{\textbf{Identifies an Impactful Vulnerability:} The paper formalizes a
    novel security flaw where a dishonest provider can bypass the intended
    resource costs of LLM serving. By demonstrating that a valid proof does not
    necessarily imply a ``Proof of Computation,'' the authors reveal a
    significant economic and security risk for verifiable outsourced inference.}

    \item \added{\textbf{Provides a Valuable Step Forward in an Established Field:} The
    work provides a necessary bridge between theoretical cryptography and
    practical machine learning security. The proposed ``ghost weight''
    constructions (Attacks A and B) are technically sound and articulately
    explained, offering a principled way to evaluate the limitations of existing
    Transformer-based zkML schemes.}
\end{enumerate}

\added{\subsection*{Noteworthy Concerns}}
\begin{enumerate}
    \item \addedd{The motivation for inflating the declared model size is weak. In many
    settings, users evaluate models based on output quality rather than
    parameter count. If the gap between claimed and actual model size is large,
    the attack may become detectable; if the gap is small, the economic benefit
    may be limited. The paper lacks a more detailed discussion of realistic
    attacker incentives and cost savings.}

\end{enumerate}

\end{document}